\documentclass[prl,aps,twocolumn,superscriptaddress]{revtex4}

\usepackage{amsmath,graphicx,epsfig}
\usepackage{epstopdf}
\usepackage{euscript}
\usepackage{amsfonts}
\usepackage{amssymb}
\usepackage{float}

\def\prn#1{{\left(#1\right)}}

\def\sbrk#1{{\left[#1\right]}}

\def\bra#1{{\langle#1|}}

\def\cg(#1,#2)(#3,#4)(#5,#6){\bra{#1,#2,#3,#4}#5,#6\rangle}

\def\ts#1{{_{\mbox{\scriptsize #1}}}}

\def\threej(#1,#2)(#3,#4)(#5,#6){\begin{pmatrix}#1&#3&#5\\#2&#4&#6\end{pmatrix}}
\def\sixj(#1,#2,#3)(#4,#5,#6){\begin{Bmatrix}#1&#2&#3\\#4&#5&#6\end{Bmatrix}}
\def\ninej(#1,#2,#3)(#4,#5,#6)(#7,#8,#9){\begin{Bmatrix}#1&#2&#3\\#4&#5&#6\\#7&#8&#9\end{Bmatrix}}

\def\mb{\mathbf}
\def\bs{\boldsymbol}

\newlength{\defbaselineskip}
\newcommand{\setlinespacing}[1]%
           {\setlength{\baselineskip}{#1 \defbaselineskip}}

\begin{document}

\title{Searching for axion stars and Q-balls with a terrestrial magnetometer network} 

\author{D. F. Jackson Kimball}
\email{derek.jacksonkimball@csueastbay.edu}
\affiliation{Department of Physics, California State University -- East Bay, Hayward, California 94542-3084, USA}

\author{D. Budker}
\affiliation{Johannes Gutenberg-Universit\"at Mainz, 55128 Mainz, Germany}
\affiliation{Helmholtz Institut Mainz, 55099 Mainz, Germany}
\affiliation{Department of Physics, University of California at Berkeley, Berkeley, California 94720-7300, USA}
\affiliation{Nuclear Science Division, Lawrence Berkeley National Laboratory, Berkeley, California 94720, USA}

\author{J. Eby}
\affiliation{Department of Physics, University of Cincinnati, Cincinnati, OH 45221, USA}
\affiliation{Fermi National Accelerator Laboratory, P.O. Box 500, Batavia, IL 60510, USA}


\author{M. Pospelov}
\affiliation{Department of Physics and Astronomy, University of Victoria, Victoria, British Columbia V8P 1A1, Canada}
\affiliation{Perimeter Institute for Theoretical Physics, Waterloo, Ontario N2J 2W9, Canada}

\author{S. Pustelny}
\affiliation{Institute of Physics, Jagiellonian University, 30-059 Krak\'ow, Poland}

\author{T. Scholtes}
\affiliation{Physics Department, University of Fribourg, CH-1700 Fribourg, Switzerland}

\author{Y. V. Stadnik}
\affiliation{Johannes Gutenberg-Universit\"at Mainz, 55128 Mainz, Germany}
\affiliation{Helmholtz Institut Mainz, 55099 Mainz, Germany}

\author{A. Weis}
\affiliation{Physics Department, University of Fribourg, CH-1700 Fribourg, Switzerland}

\author{A. Wickenbrock}
\affiliation{Johannes Gutenberg-Universit\"at Mainz, 55128 Mainz, Germany}

\date{\today}



\begin{abstract}
Light (pseudo-)scalar fields are promising candidates to be the dark matter in the Universe. Under certain initial conditions in the early Universe and/or with certain types of self-interactions, they can form compact dark-matter objects such as axion stars or Q-balls. Direct encounters with such objects can be searched for by using a global network of atomic magnetometers. It is shown that for a range of masses and radii not ruled out by existing observations, the terrestrial encounter rate with axion stars or Q-balls can be sufficiently high (at least once per year) for a detection. Furthermore, it is shown that a global network of atomic magnetometers is sufficiently sensitive to pseudoscalar couplings to atomic spins so that a transit through an axion star or Q-ball could be detected over a broad range of unexplored parameter space.
\end{abstract}



\maketitle

A host of astrophysical and cosmological measurements suggest that over 80\% of all matter in the Universe is dark matter \cite{Ber05,Fen10,Gor14}. In order to elucidate the nature of dark matter, terrestrial experiments seek to measure non-gravitational interactions of dark matter with standard-model particles and fields. However, extrapolation from the $\gtrsim 1~{\rm kpc}$ distances associated with astrophysical observations to particle-physics phenomena accessible to laboratory-scale experiments leaves open a vast number of plausible theoretical possibilities worth exploring.

A well-motivated hypothesis is that a substantial fraction of dark matter consists of ultralight bosons such as axions \cite{Pre83,Abb83,Din83} or axion-like particles (ALPs \cite{Svr06,Arv10,Gra15}) with masses $m_a c^2 \lesssim 10~{\rm eV}$.  Such ultralight bosons will have a large number density in the galaxy and thus their phenomenology is well-described by a classical field. In this scenario, the mass-energy associated with dark matter is primarily stored in coherent oscillations of the dark-matter field \cite{Duf09,Gra13,Sta14,Bud14}. There are a number of proposed and ongoing searches for continuous oscillatory signals generated by the axion/ALP dark matter background assuming that terrestrial detectors are bathed in a continuous dark-matter flux, see for example Refs.~\cite{Bud14,Asz01,Asz10,Sik14,Kah16,Abe17,Cal17MADMAX}.

However, it is possible that instead of a roughly uniform distribution throughout the halo, ALPs could be concentrated in compact objects. For example, the mass-energy associated with the Universe's dark sector (dark matter and partially dark energy) could be stored primarily in topological defects such as domain walls, strings or monopoles \cite{Vil85,Bat99,Pos13}. Another plausible scenario is that initial inhomogeneities in the galactic dark matter distribution enable gravity or self-interactions \cite{Spe00} to generate bound clumps or ``stars'' composed of ALPs \cite{Kau68,Ruf69,Bre84,Tka91,Kol93,Kol94,Sch03,Bar11,Bar13,Lie12,Eby15,Gut15,Bra16,Kli17}. A prominent, closely related example of a compact, composite dark-matter object is the Q-ball \cite{Ros68,Col85,Lee87,Lee92,Kus01} or Q-star, a non-topological soliton of a light scalar field \cite{Lee87,Lee92}. In this work, we are primarily interested in ALP stars or Q-stars (collectively referred to as soliton stars \cite{Lee87}) with radii $R \gg R_E$ (the radius of Earth).  Under conditions where the attractive interactions between ALPs are sufficiently strong so that most of the dark-matter mass takes the form of soliton stars, instead of being bathed in a continuous dark-matter flux, terrestrial detectors will instead witness transient events when Earth passes through the soliton stars \cite{endnote1}.

Dark-matter fields consisting of ALPs are generically predicted to interact, albeit feebly, with the intrinsic spins of elementary particles \cite{Moo84,Dob06}. The {\textbf{G}}lobal {\textbf{N}}etwork of {\textbf{O}}ptical {\textbf{M}}agnetometers to search for {\textbf{E}}xotic physics (GNOME) collaboration \cite{Pos13,Pus13} is presently conducting a search for transient spin-dependent interactions that might arise, for example, if Earth passes through a compact dark-matter object. While a single atomic-magnetometer system could in principle detect such transient events, in practice it is difficult to confidently distinguish a true signal heralding new physics from ``false positives'' induced by occasional abrupt changes of magnetometer operational conditions. To veto false positive events, suppress noise, and effectively characterize true exotic transient signals, the GNOME consists of an array of magnetometers widely distributed over Earth's surface. Crucially, the geographically distributed array of magnetometers enables consistency checks based on the relative timing and amplitudes of transient signals, suppressing the stochastic background. Data analysis is based on proven techniques developed by the Laser Interferometer Gravitational Wave Observatory (LIGO) collaboration \cite{And01,All12} to search for similar correlated ``burst'' signals from a worldwide network of gravitational-wave detectors. It was demonstrated that these techniques can be adapted to analyze GNOME data in Ref.~\cite{Pus13}.

A complementary experimental approach is being pursued to search for massive compact dark-matter objects using atomic clocks as sensors rather than atomic magnetometers \cite{Der14}. While atomic magnetometers are sensitive to fields that couple to spins (such as the pseudoscalar interaction associated with ALPs \cite{Moo84,Dob06}), atomic clocks are sensitive to fields that effectively alter the values of fundamental constants, such as the fine-structure constant, through scalar interactions. An encounter with a soliton star that has such scalar interactions would manifest as a ``glitch'' propagating through an atomic-clock network, such as the Global Positioning System (GPS) \cite{Der14}: clocks would become desynchronized as Earth passes through the soliton star. The GPS.DM collaboration is analyzing data from the GPS satellites and have recently produced the first constraints on such events \cite{Rob17}. There have also been recent proposals to search for transient signals generated by exotic physics using networks of laser/maser interferometers \cite{Sta15,Sta16}, resonant-bar detectors \cite{Jac15,Arv16,Bra17}, and pulsar timing \cite{Sta14pulsar}.

Here we analyze the prospects for observing a transient signal from an encounter with a soliton star using a terrestrial detector network, with the GNOME as a concrete example. Presently, the GNOME consists of six dedicated optically pumped atomic magnetometers \cite{Bud13} located at stations throughout the world (nine additional new stations are under construction \cite{GNOMEwebsite}). The magnetometric sensitivity of existing GNOME sensors is $\delta B \approx 100~{\rm fT/\sqrt{Hz}}$ over a bandwidth of $\approx 100~{\rm Hz}$. The GNOME is primarily sensitive to exotic interactions of electrons and protons \cite{Kim15}. A next-generation Advanced GNOME is under development that will use alkali-$^3$He comagnetometers \cite{Kor02,Kor05,Vas09,Bro10} and will be primarily sensitive to neutron interactions \cite{Kim15}. Advanced GNOME sensors will have effective sensitivities of $\delta B \approx 1~{\rm fT/\sqrt{Hz}}$ to ``pseudo-magnetic fields'' caused by ALP interactions over a similar bandwidth \cite{Kor02,Kor05}. The GNOME magnetometers are located within multi-layer magnetic shields to reduce the influence of external magnetic noise and perturbations, while still maintaining sensitivity to exotic spin-dependent interactions \cite{Kim16}. Each GNOME sensor also uses auxiliary unshielded magnetometers and sensors, such as accelerometers and gyroscopes, to measure relevant environmental conditions, enabling the exclusion of data with known systematic issues. The signals from the GNOME sensors are recorded with accurate timing using a custom GPS-disciplined data acquisition system \cite{Wlo14} and have a characteristic temporal resolution of $\lesssim 10~{\rm ms}$ (determined by the magnetometer bandwidths).

One of the most important questions at the outset of our considerations is whether it is theoretically plausible that Earth would encounter a soliton star over the course of an observational period of $\sim 1$~year. Certainly (and fortunately!), stars composed of ordinary matter are so dilute within our galaxy that collisions are extraordinarily infrequent on human time scales, and one might be concerned whether this is also true of soliton stars for any reasonable parameters. We also consider whether encounters with soliton stars having the requisite characteristics for detection by the GNOME or other similar terrestrial detector networks might be ruled out by other observations (e.g., the stability of lunar and planetary orbits in our solar system, lunar laser ranging, gravimeter data, gravitational microlensing studies, etc.). Finally, we investigate the parameter space over which the GNOME is sensitive to soliton-star transits given the GNOME's technical characteristics (sensitivity, bandwidth, etc.).

To determine the soliton-star parameter space to which a terrestrial detector network is sensitive, we begin by assuming a uniform local distribution of soliton stars. The characteristic relative velocity of our solar system with respect to other objects in the galaxy is given by the local virial velocity $v \sim 10^{-3}c$. Thus in order for soliton stars to be detectable with the GNOME during a $1$-year observational period, the mean-free-path length $L$ between soliton stars must be $\lesssim L\ts{max} = 10^{-3}~{\rm ly}$, where
\begin{align}
L \approx \frac{1}{n \pi R^2} \approx \frac{Mc^2}{\rho\ts{DM} \pi R^2} \lesssim L\ts{max}~.
\label{Eq:mean-free-path}
\end{align}
Here $n$ is the number density of soliton stars, $R$ is the characteristic soliton-star radius, the local dark-matter energy density is $\rho\ts{DM} \approx 0.4~{\rm GeV/cm^3}$ \cite{Sof01,Jun96,Ber98,Cat12}, and $M$ is the characteristic soliton-star mass. We assume that the bulk of the dark matter is in the form of soliton stars, so that $\rho\ts{DM} \approx n M c^2$.  This establishes an upper limit on $R$ since the concept of a compact dark-matter object only makes sense if $R \ll L\ts{max} \approx 10^6 R_E$, where $R_E$ is Earth's radius. In turn, this gives an upper limit on the soliton star mass based on Eq.~\eqref{Eq:mean-free-path}, $Mc^2 \ll 10^{54}~{\rm eV} \approx 10^{-12} M_\odot$, where $M_\odot$ is the mass of the Sun.

Do existing observations rule out frequent encounters with soliton stars of this size? Searches for gravitational microlensing due to MAssive Compact Halo Objects (MACHOs) constrain their masses to be $\lesssim 10^{-7} M_\odot$ \cite{Her04}. It was shown that objects of any size with masses $\lesssim 10^{-10} M_\odot$ would not create measurable consequences for the Solar system or Earth-Moon dynamics \cite{Gon13}. Recent limits on primordial black holes from gravitational femtolensing (light deflection of $\sim 10^{-15}$ arcseconds \cite{Gou92}) of gamma-ray bursts show that objects with masses in the range $10^{-16} M_\odot$ to $10^{-13} M_\odot$ do not compose a dominant fraction of dark matter \cite{Bar12}. The gravitational femtolensing constraint from gamma-ray bursts rules out the most massive soliton stars considered above, so the observationally allowed range of soliton-star masses that could be encountered during a one year search is
\begin{align}
M \lesssim 10^{-16} M_\odot~.
\label{Eq:mass-range}
\end{align}
To simplify the analysis of the GNOME data, we assume that the size of an encountered dark matter object is much larger than Earth's radius, i.e., $R \gg R_E$: in this case, \emph{all} GNOME sensors would register a transient signal within the time $T \sim 2R_E/v$ it takes for Earth to pass through the surface of the dark-matter object. For concreteness, we assume $R\ts{min} \sim 10 R_E$, and thus for the soliton star radii we have
\begin{align}
10R_E \lesssim R \lesssim 10^6 R_E~.
\label{Eq:radius-range}
\end{align}
For context, the most massive soliton stars considered here correspond to the average mass of a comet.

In principle, the acceleration due to the gravitational force from an encounter with a soliton star offers another avenue for detection. However, the peak acceleration felt during an encounter would be
\begin{align}
g_a \approx \frac{GM}{R^2} \approx \frac{\pi G \rho\ts{DM} L\ts{max}}{c^2} \approx 3 \times 10^{-16}~{\rm cm/s^2}~,
\label{Eq:acceleration}
\end{align}
or $3 \times 10^{-19}g$ (where $g \approx 10^3~{\rm cm/s^2}$ is the acceleration due to Earth's gravity). This is far smaller than even the best accelerometers could conceivably measure \cite{Hu13}. The tidal effects of such a soliton-star encounter on gravitational-wave observatories such as LIGO are orders of magnitude below LIGO's strain sensitivity, and in any case would tend to be excluded from detection by LIGO because they would not generate tensor effects on the two interferometer arms. Moreover, the inverse time of the passage, $10^{-3}c/R \lesssim 10^{-2}\,{\rm Hz}$, is in the frequency domain where LIGO has poor sensitivity due to seismic noise.

Having established that sufficiently frequent encounters with soliton stars are both possible and not ruled out by existing observations, the next question is whether the GNOME has sufficient sensitivity to detect a transient event resulting from a terrestrial encounter with such a soliton star. At this point, we adopt a more specific theoretical model for estimates, namely the Q-star model of Refs.~\cite{Col85,Lee87,Lee92,Kus01}, although it turns out that our conclusions are quite generic for soliton stars of the considered sizes regardless of the details of the attractive interactions holding the ALPs together (so long as they are sufficiently strong). In the model described in Refs.~\cite{Col85,Lee87,Lee92,Kus01}, the Q-stars arise as a consequence of a particle-antiparticle asymmetry of a complex scalar field and its self-interaction. Consider a region of space where a complex scalar field oscillates at angular frequency $\omega$ (not necessarily the Compton frequency),
\begin{align}
a\prn{ \mb{r},t } = e^{i \omega t}\phi(\mb{r})~.
\label{Eq:oscillating-ALP-field}
\end{align}
Such a configuration possesses a conserved additive quantum number $Q$ (where each individual ALP has charge $Q=1$), in which case \cite{Col85}
\begin{align}
Q = \frac{\omega}{\hbar^2c^3} \int \left| \phi(\mb{r}) \right|^2 d^3\mb{r}~.
\label{Eq:Q-definition}
\end{align}
[From now on, we shall refer to $a\prn{ \mb{r},t }$ as a generalized ALP field.] The necessary conditions for the appearance of Q-stars are that $Q \neq 0$ averaged over the whole space, and a self-interaction potential $U(\phi)$ possessing at least two distinct minima at $\phi = 0$ and at $\phi = \phi_0$ \cite{Col85,Lee87,Lee92,Kus01}.
If initially there exist regions of space with different energy vacua, regions where $\phi = \phi_0$ can deform but not disappear entirely because of the conserved charge $Q$.
Furthermore, $U(\phi)$ is nonzero in the Q-star's transitional surface region where $\phi$ goes from $\phi_0$ to $0$ \cite{Lee87,Lee92}; $U(\phi_0)$ in the interior of the Q-star may also be nonzero \cite{Col85}, but this is not required \cite{Lee87,Lee92} and so for simplicity we set $U(\phi_0) = 0$ here. Thus the Q-star possesses a potential energy per unit surface area of $\sigma$, where $\sigma$ is a constant depending on the particular properties of $U(\phi)$ \cite{endnote2}. The total energy of a Q-star with volume $V$ and surface area $A$ is
\begin{align}
E = \hbar \omega Q + \sigma A~,
\label{Eq:Q-star-energy-1}
\end{align}
where each ALP within the Q-star contributes a quantum of energy $\hbar \omega$. To minimize the energy of the field configuration while conserving the charge $Q$, we express $\omega$ in terms of $Q$ using Eq.~\eqref{Eq:Q-definition},
\begin{align}
\omega = \hbar^2 c^3 \frac{Q}{\phi_0^2V}~,
\label{Eq:frequency-in-Q-star}
\end{align}
and thus
\begin{align}
E = \hbar^3c^3 \frac{Q^2}{\phi_0^2 V} + \sigma A~.
\label{Eq:Q-star-energy-2}
\end{align}
Thus the energy is minimized when the Q-star assumes a spherical shape which minimizes $A$ and maximizes $V$. Minimizing $E$ with respect to $R$, one arrives at the total mass-energy of the Q-star:
\begin{align}
E = Mc^2 = \frac{5}{2} \hbar \omega Q = \frac{10\pi}{3 \hbar c^3} \omega^2 \phi_0^2 R^3~,
\label{Eq:Q-star-mass}
\end{align}
where in the last step we have substituted the relationships from Eqs.~\eqref{Eq:Q-definition} and \eqref{Eq:frequency-in-Q-star}.

Note that it is energetically favorable for ALPs to remain within the Q-star if $\omega \lesssim \omega_a$, where $\omega_a$ is the ALP Compton frequency, since ALPs inside the Q-star have energy $\hbar\omega$ while those outside the Q-star have energy $\hbar\omega_a$. The values of $\omega^2$ and $\omega_a^2$ are proportional to $\partial^2 U / \partial \phi^2$ at the respective potential minima inside ($\phi=\phi_0$) and outside ($\phi=0$) the Q-star and can thus be different \cite{Col85,Kus01}.  The condition $\omega \lesssim \omega_a$ ensures stability of the Q-star with respect to radiative decay via ALP emission. In the described scenario, the oscillating ALP field exists only within the Q-stars and thus evades detection by terrestrial experiments searching for a uniform dark-matter field.

The GNOME is sensitive to encounters with such Q-stars (and axion/ALP stars in general) through the coupling of the ALP field to the intrinsic spins of standard model fermions. The gradient of a real-valued ALP field can couple to the spin $\mb{S}_i$ of a particle $i$ through a nonrelativistic Hamiltonian (the so-called linear interaction)
\begin{align}
H\ts{lin,\it{i}} = \frac{\hbar c}{f\ts{lin,\it{i}}} \mb{S}_i \cdot {\bs{\nabla}} a~.
\label{Eq:lin-ALP-Hamiltonian}
\end{align}
Here $\mb{S}_i$ is in units of $\hbar$ and $f\ts{lin,\it{i}}$ (having dimensions of energy) is related to the coupling constant for the considered particle $i$, and can be different for electrons, neutrons, and protons \cite{Pos13}. We treat the coupling constant $f\ts{lin,\it{i}}$, apart from experimental and observational limits, as a free parameter.

In a theory with one real-valued ALP field, interactions with standard model fermions result from the Lagrangian density given by the coupling of the
space-time derivative of the ALP field $a$ to fermion axial-vector currents,
\begin{align}
{\cal L} \propto \frac{1}{f\ts{lin,\it{i}}}\partial_\mu a \times \bar \psi_{i} \gamma_\mu\gamma_5 \psi_{i}~,
\end{align}
where $\psi_i$ represents the fermion field and $\gamma_\mu$ and $\gamma_5$ are Dirac matrices. For a complex-valued field $a$ forming Q-stars, such a form of ${\cal L}$ is inconsistent with the $U(1)_Q$ symmetry in the $a$ sector. In that case, the interactions would have to be bilinear in $a$. A possible form of such an interaction consistent with $U(1)_Q$ can then be
\begin{align}
{\cal L} \propto \frac{1}{(f\ts{quad,\it{i}})^2}\partial_\mu (a^*a) \times \bar \psi_{i} \gamma_\mu\gamma_5 \psi_{i}~,
\label{quad1}
\end{align}
or alternatively
\begin{align}
{\cal L} \propto \frac{1}{(f\ts{quad,\it{i}})^2}i\sbrk{a^*\partial_\mu a-(\partial_\mu a^*)a} \times \bar \psi_{i} \gamma_\mu\gamma_5 \psi_{i}.
\label{quad2}
\end{align}
The nonrelativistic Hamiltonian corresponding to the second case [Eq.~\eqref{quad2}] is proportional to the gradient of the square of the field (the so-called quadratic interaction):
\begin{align}
H\ts{quad,\it{i}} = \frac{\hbar c}{\prn{f\ts{quad,\it{i}}}^2} \mb{S}_i \cdot i\sbrk{a^*{\bs{\nabla}} a - ({\bs{\nabla}} a^*)a}~,
\label{Eq:quad-ALP-Hamiltonian}
\end{align}
while for the first case [Eq.~\eqref{quad1}] the corresponding combination of scalar fields is ${\bs{\nabla}} |a|^2$.

Astrophysical observations disfavor ALPs with nucleon couplings $f\ts{lin,\it{np}} \lesssim 10^9~{\rm GeV}$ \cite{Raf99} and electron couplings $f\ts{lin,\it{e}} \lesssim 10^{10}~{\rm GeV}$ \cite{Raf08,Cor01}. Astrophysical constraints on the quadratic interaction are less stringent: so far ALPs with $f\ts{quad,\it{i}} \lesssim 10^4~{\rm GeV}$ are disfavored \cite{Oli08,Pos13}.

In order to understand the effect of such an interaction on atomic spins during the transit of Earth through a soliton star, we need to understand the behavior of ${\bs{\nabla}} a$ during the transit. In principle, there are two components to such a gradient term: the first one is related to a gradient of the ``envelope'' $\phi(\bf{r})$ and it would exist even in the limit of vanishing relative velocity. The second effect is due to a combination of the time-dependence of $a(\mb{r},t)$ [Eq.~\eqref{Eq:oscillating-ALP-field}]
and a nonzero relative velocity between the soliton star and the detector. Based on Eqs.~\eqref{Eq:mean-free-path} and \eqref{Eq:Q-star-mass}, we can approximate the ALP field amplitude $\phi_0$ as a step-function with a value inside the Q-star given by:
\begin{align}
\phi_0^2 \approx \frac{3\hbar c^3}{10\pi} \frac{\rho\ts{DM} L}{\omega^2R}~.
\end{align}
For the time being, we neglect terms associated with ${\bs{\nabla}} \phi(\bf{r})$, while the relative motion creates a nonzero oscillating spin-dependent energy shift. The amplitude of the oscillating gradient of $a$ is given by
\begin{equation}
\left|\bs{\nabla} a \right| \approx \frac{\omega v}{c^2} \phi_0~,
\label{Eq:axion-gradient}
\end{equation}
where $v \sim 10^{-3}c$ is the relative velocity between the soliton star and the terrestrial detectors, and similarly
\begin{equation}
\left|a^*{\bs{\nabla}} a - ({\bs{\nabla}} a^*)a \right| \approx \frac{2\omega v}{c^2} \phi_0^2~.
\label{Eq:axion-gradient}
\end{equation}
Thus we estimate that for the linear coupling to spins given by Eq.~\eqref{Eq:lin-ALP-Hamiltonian}, atomic spins will experience an oscillating energy shift inside an ALP Q-star of amplitude
\begin{align}
\Delta E\ts{lin,\it{i}} \approx \frac{1}{f\ts{lin,\it{i}}} \frac{v}{c} \sqrt{ \frac{3\hbar^3 c^3 \rho\ts{DM} L}{10 \pi R} }~.
\label{Eq:lin-ALP-energy-shift}
\end{align}
For the quadratic coupling to spins [Eq.~\eqref{Eq:quad-ALP-Hamiltonian}],
\begin{align}
\Delta E\ts{quad,\it{i}} \approx \frac{1}{\prn{f\ts{quad,\it{i}}}^2} \frac{ v}{\omega c}  \frac{3\hbar^2c^3}{5\pi} \frac{\rho\ts{DM} L}{R}~.
\label{Eq:quad-ALP-energy-shift}
\end{align}
Equations \eqref{Eq:lin-ALP-energy-shift} and \eqref{Eq:quad-ALP-energy-shift} can be used to translate the sensitivity to spin-dependent energy shifts of a GNOME magnetometer into a sensitivity to the coupling constants $f\ts{lin,\it{i}}$ and $f\ts{quad,\it{i}}$. It is important to note that the energy-shift sensitivity depends not only on the magnetic-field sensitivity but also on the coupling/particle probed; in particular, for a given magnetometric sensitivity, a noble-gas magnetometer is more sensitive to energy shifts than an alkali-atom-based magnetometer because of the difference between the Bohr and nuclear magnetons.

In order to estimate the parameter space accessible to GNOME, we conservatively take $\omega \sim \omega_a = m_a c^2/\hbar$, since $\omega \lesssim \omega_a$ and smaller values of $\omega$ lead to larger energy shifts according to Eq.~\eqref{Eq:quad-ALP-energy-shift}. The sensitivity of the GNOME to a soliton-star encounter is not only determined by $\phi_0$, but also by the characteristic frequency and duration of a signal. Because ALPs cannot be confined to a region smaller than their Compton wavelength $\lambda_a$, this demands that $R \gtrsim \lambda_a$ \cite{Eby16b,Cha11a,Cha11b}; the minimum detectable soliton-star radius of $R\ts{min} \approx 10R_E$ corresponds to Compton frequencies $\omega_a/(2\pi) \approx 1~{\rm Hz}$. Below ALP masses $m_a$ corresponding to $\omega_a \approx 2\pi \times {\rm 1~Hz}$, the minimum radius of a soliton star is $\lambda_a$ rather than $10R_E$. The upper limit on the detectable $m_a$ is set by the GNOME bandwidth of $\approx 100~{\rm Hz}$.

\begin{figure}
\center
\includegraphics[width=3.25 in]{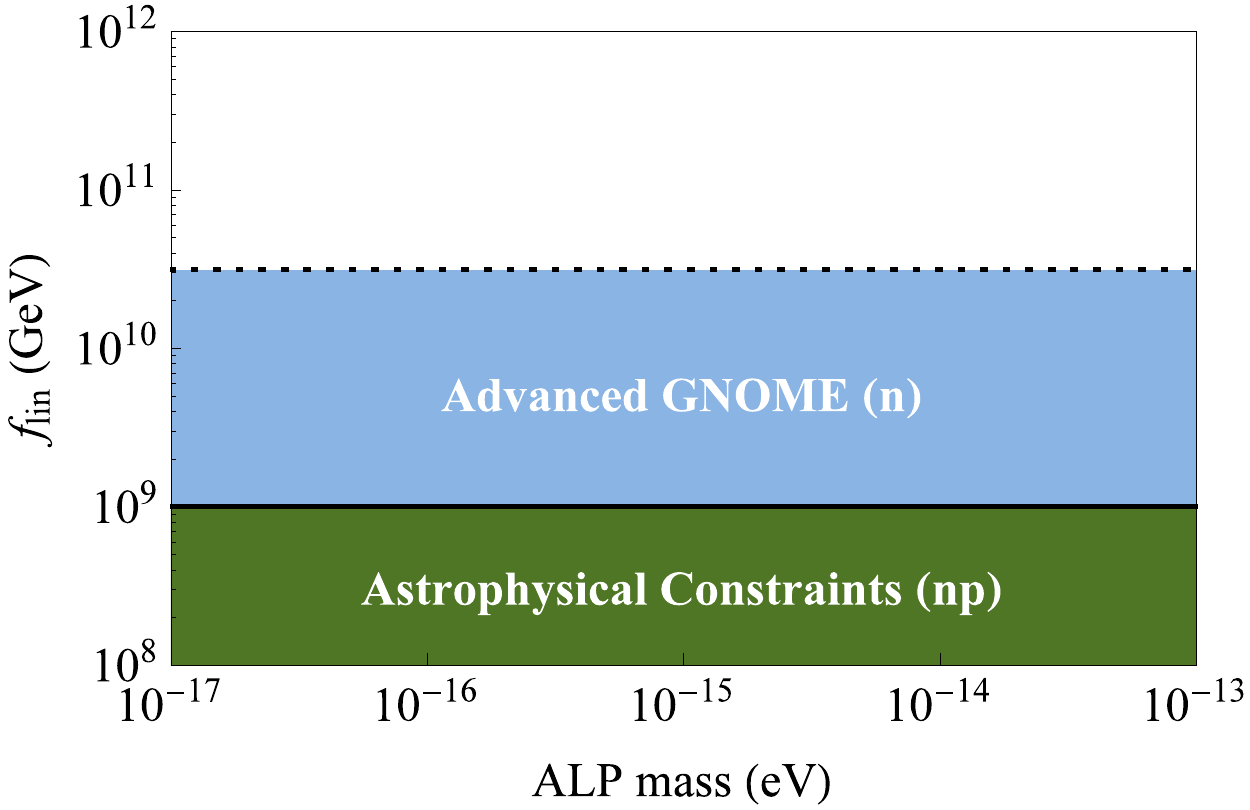}
\caption{Estimated parameter space probed by the Advanced GNOME (dotted line, light blue fill) for the linear interaction of neutron spins with an ALP star, assuming that the mean-free-path length for terrestrial encounters with ALP stars is $L = 10^{-3}~{\rm ly}$ and $v = 10^{-3} c$. The solid line and green fill represent existing astrophysical constraints on spin-dependent ALP interactions with nucleons \cite{Raf99}. The sensitivity of the existing GNOME is slightly below the level of the astrophysical constraints.}
\label{Fig:ALPstar-GNOME-lin}
\end{figure}

\begin{figure}
\center
\includegraphics[width=3.25 in]{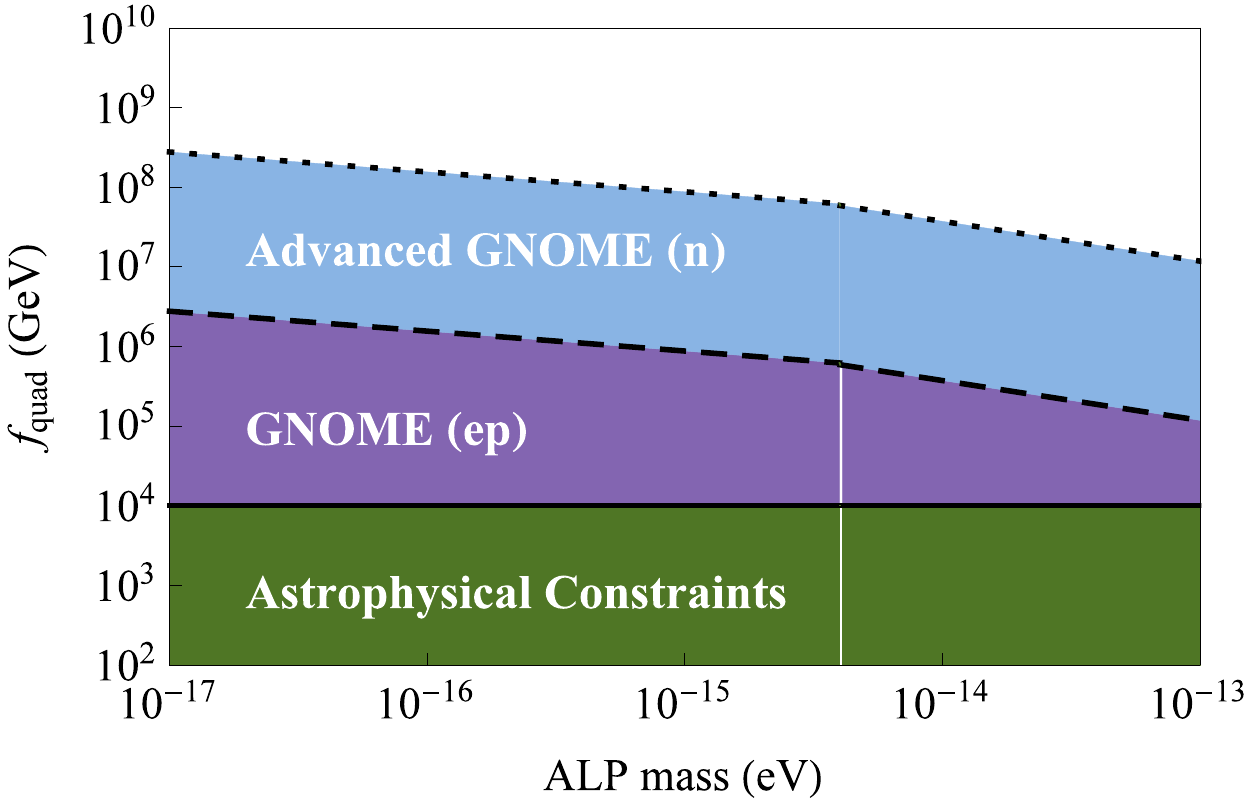}
\caption{Estimated parameter space probed by the existing GNOME (dashed line, purple fill) and by the Advanced GNOME (dotted line, light blue fill) for the quadratic interaction of electron/proton spins or neutron spins, respectively, with an ALP star \cite{Kim15}. We assume that $L = 10^{-3}~{\rm ly}$, and for $m_a > 4 \times 10^{-15}~{\rm eV}$, $R=R\ts{min}=10R_E$, and for $m_a < 4 \times 10^{-15}~{\rm eV}$, $R=\lambda_a$. The solid line and green fill represent existing astrophysical constraints \cite{Oli08,Pos13}.}
\label{Fig:ALPstar-GNOME-quad}
\end{figure}

The energy resolution of a GNOME magnetometer for a given transient event depends on the duration of the event $\tau$, which determines the signal integration time. The duration of a soliton-star encounter is $\tau \sim R/v$ (which corresponds to $\approx 200~{\rm s}$ for $R\ts{min} \approx 10R_E$),
\begin{align}
\Delta E \approx \frac{\hbar \gamma_i \delta B}{ \sqrt{ \tau } } \approx \hbar \gamma_i \delta B \sqrt{ \frac{v}{R} }~,
\label{Eq:GNOME-sensitivity}
\end{align}
where $\gamma_i$ is the gyromagnetic ratio for particle $i$ and $\delta B$ is the magnetometric sensitivity per root Hz. Comparing Eq.~\eqref{Eq:GNOME-sensitivity} to Eqs.~\eqref{Eq:lin-ALP-energy-shift} and \eqref{Eq:quad-ALP-energy-shift}, we find the sensitivity of GNOME to the coupling constants $f\ts{lin,\it{i}}$ and $f\ts{quad,\it{i}}$ to be:
\begin{align}
\Delta f\ts{lin,\it{i}} \approx \frac{1}{\gamma_i \delta B} \sqrt{ \frac{ 3 v \hbar c \rho\ts{DM} L}{10\pi} }~,
\label{Eq:GNOME-sensitivity-linear}
\end{align}
and
\begin{align}
\Delta \prn{f\ts{quad,\it{i}}}^2 \approx \frac{1}{\gamma_i \delta B} \frac{3\hbar^2 \rho\ts{DM} L}{5\pi m_a} \sqrt{ \frac{v}{R} }~.
\label{Eq:GNOME-sensitivity-quad}
\end{align}
The parameter space of spin couplings that can be probed during an ALP star encounter by the GNOME and the Advanced GNOME is shown in Figs.~\ref{Fig:ALPstar-GNOME-lin} and \ref{Fig:ALPstar-GNOME-quad}, assuming $L = 10^{-3}~{\rm ly}$, $v = 10^{-3} c$, and $R = 10R_E$. This not only shows that it is possible for the GNOME to detect an ALP star encounter given existing constraints on ALP couplings, but also that the GNOME is sensitive to many decades of unexplored parameter space.

\begin{figure}
\center
\includegraphics[width=3.25 in]{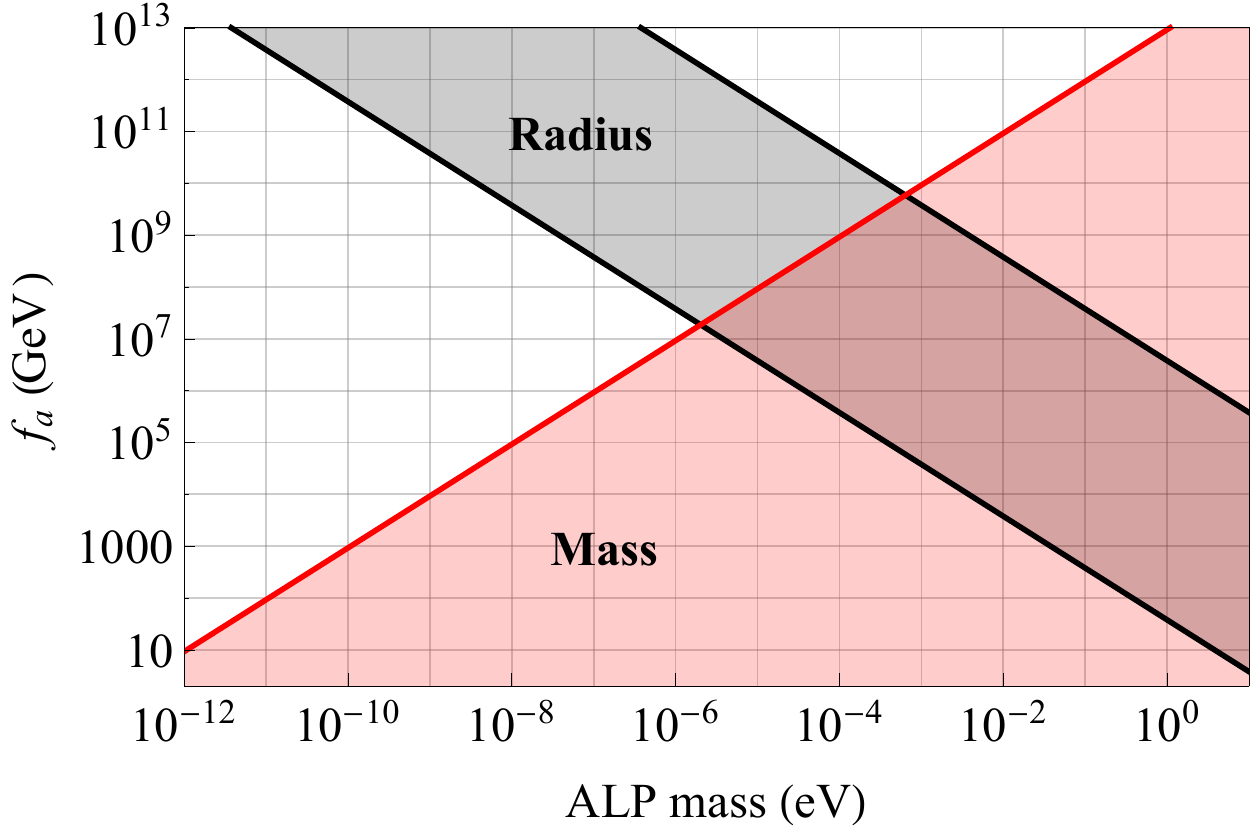}
\caption{Parameter space for which ALP stars assume typical sizes consistent with relatively frequent encounters with Earth (on time scales $\lesssim 1~{\rm yr}$) and are not ruled out by other observations, based on the specific model for the ALP potential $U(\phi)$ given by Eq.~\eqref{Eq:CosModel-potential}. The gray band bounded by solid black lines shows the range of $f_a$ and $m_a$ values consistent with typical ALP star radii $R$ in the range $10R_E \lesssim R \lesssim 10^6 R_E$ [Eq.~\eqref{Eq:radius-range}]. The pink area bounded by a solid red line shows the range of $f_a$ and $m_a$ values consistent with typical ALP star masses $M \lesssim 10^{-16} M_\odot$ [Eq.~\eqref{Eq:mass-range}]. The region of parameter space where the two shaded areas overlap shows the values of $f_a$ and $m_a$ consistent with both constraints.}
\label{Fig:AvgALPstar}
\end{figure}

The scenario in which the GNOME could detect a terrestrial transit through a soliton star hinges upon the average soliton-star size being within the particular range identified by Eqs.~\eqref{Eq:mass-range} and \eqref{Eq:radius-range} where such transits are sufficiently frequent. Under what conditions of creation might the average soliton-star size fall within this range? Previous studies concerning the production and evolution of Q-stars in the early universe, for instance, have found their sizes and masses to be model-dependent and generally unconstrained \cite{Gri89,Fri89,Kus97,Kus98}. On the other hand, one can explore the plausibility of this scenario in more detail by investigating a specific model for the ALP interaction potential $U(\phi)$. For example, by employing the axion-star model discussed in Refs.~\cite{Kol93,Bar11,Bar13,Gut15,Bra16}, the average size and mass of soliton stars can be related to parameters describing $U(\phi)$ (such issues have also been explored, for example, in Refs.~\cite{Kol94clumps,Cha11a,Cha11b,Eby15,Ena17axion}). The model of Refs.~\cite{Kol93,Bar11,Bar13,Eby15,Gut15,Bra16} assumes the potential
\begin{align}
U(\phi) = \frac{c}{\hbar^3}f_a^2 m_a^2 \sbrk{ 1 - \cos\prn{ \frac{\phi}{f_a} } }~,
\label{Eq:CosModel-potential}
\end{align}
where $f_a$ is the ALP decay constant. Typically, $f_a \sim f\ts{lin}$ (or $f\ts{quad}$), although this can be model dependent \cite{Raf99}. As shown in Ref.~\cite{Eby15}, for example, the radius and mass of such a soliton star scale as
\begin{align}
R \sim \frac{\hbar c}{f_a}\frac{M\ts{Pl}}{m_a}~,
\label{Eq:CosModel-R}
\end{align}
\begin{align}
M \sim \frac{M\ts{Pl} f_a}{m_ac^2}~,
\label{Eq:CosModel-M}
\end{align}
where $M\ts{Pl} = \sqrt{ \hbar c / G }$ is the Planck mass. The average radius and mass of a soliton star should be similar to that described by Eqs.~\eqref{Eq:CosModel-R} and \eqref{Eq:CosModel-M} \cite{Bar11,Bar13,Eby15,Bra16}. In this scenario, the range of average soliton star masses and radii for which terrestrial encounters are sufficiently frequent [Eqs.~\eqref{Eq:mass-range} and \eqref{Eq:radius-range}] determines a corresponding range of $m_a$ and $f_a$, as shown in Fig.~\ref{Fig:AvgALPstar}. This further demonstrates that the detection of ALP star encounters with a terrestrial detector network is feasible in some scenarios.

The range of possible $f_a$ identified in Fig.~\ref{Fig:AvgALPstar} for the specific $U(\phi)$ of Eq.~\eqref{Eq:CosModel-potential} goes up to $f_a \sim 10^{10}~{\rm GeV}$, which is beyond existing astrophysical constraints on $f\ts{lin}$ and $f\ts{quad}$ as shown in Fig.~\ref{Fig:ALPstar-GNOME-lin} and \ref{Fig:ALPstar-GNOME-quad} and thus not ruled out. Although the accessible range of $m_a$ ($m_ac^2 \gtrsim 10^{-6}~{\rm eV}$) corresponds to ALP-field oscillation frequencies outside the bandwidth of the GNOME in this particular model, the GNOME is sensitive, for example, to the spatial gradient of the more slowly varying envelope function $\phi({\mb{r}})^2$ for the quadratic interaction. Furthermore, high-frequency detectors such as those to be employed in the Cosmic Axion Spin Precession Experiment (CASPEr) \cite{Bud14} are sensitive at the lower end of this ALP mass range ($m_a \sim 10^{-6}~{\rm eV}$). Experiments searching for ALP-photon couplings \cite{Bra03}, such as the Axion Dark Matter eXperiment (ADMX), are sensitive to ALPs with $10^{-6}~{\rm eV} \lesssim m_ac^2 \lesssim 10^{-4}~{\rm eV}$ \cite{Asz01,Asz10,Bru17}.

In conclusion, we have shown that a terrestrial network of magnetometers such as the GNOME \cite{Pos13,Pus13} is sensitive to encounters with dark-matter stars composed of axion-like particles over many decades of theoretically plausible unexplored parameter space. In order to further evaluate the plausibility of this scenario and develop more detailed descriptions of signatures and event rates, future work will involve more detailed modeling of soliton-star creation and dynamics for the range of sizes to which GNOME is sensitive. For example, an interesting question is the role of collisions \cite{Moo00} between soliton stars on the soliton-star population dynamics in the galaxy; such soliton-star dynamics are also found to depend on the specific nature of ALP interactions \cite{Kus01,Axe00,Bat00,Bar13,Fai17,Cha17}.

\acknowledgments

The authors are sincerely grateful to Andrei Derevianko, Konstantin Zioutas, Jason Stalnaker, and Chris Pankow for useful discussions. DFJK acknowledges the support of the National Science Foundation under grant PHY-1707875. DFJK, DB, and A. Wickenbrock acknowledge the support of the Simons and Heising-Simons Foundations. YVS was supported by the Humboldt Research Fellowship. The work of JE was supported by the U.S. Department of Energy, Office of Science, Office of Workforce Development for Teachers and Scientists, Office of Science Graduate Student Research (SCGSR) program; the SCGSR program is administered by the Oak Ridge Institute for Science and Education for the DOE under contract number de-sc0014664. DB acknowledges the support of the European Research Council and DFG Koselleck. MP is supported in part by NSERC, Canada, and research at the Perimeter Institute is supported in part by the Government of Canada through NSERC and by the Province of Ontario through MEDT. SP acknowledges the support of the Polish National Science Centre within the Opus program.

\bibliography{ALP-stars-GNOME}

\begin{thebibliography}{98}
\bibitem[{end({\natexlab{a}})}]{endnote1}
\bibinfo{note}{Note that if the attractive interaction between the ALPs is not
  strong enough, soliton stars will experience gravitational shearing upon
  encountering ordinary stars. In this case, the ALP dark matter would tend to
  form streams and fragment into a quasi-uniform background rather than remain
  bound as a star-like object \cite{Pat13,Tin16,Zio17}.}
\bibitem[{end({\natexlab{b}})}]{endnote2}
\bibinfo{note}{The width of the Q-star's transitional surface region may also
  depend on the particular properties of $U(\phi)$, see related discussion in
  Refs.~\cite{Eby16b,Cha11a,Cha11b}.}
\end{thebibliography}

\end{document}